\documentstyle[amssymb,12pt]{article}

\input{tcilatex}
\begin{document}

\title{Why can an electron mass vary from zero to infinity?}
\author{Guang -jiong Ni$^{*},$Weimin Zhou$^{*}$ and Jun Yan$^{*}$ \\
$^{*}$Departmint of Physics, Fudan University\\
Shanghai 200433, P.R.China}
\maketitle

\begin{abstract}
When a particle is in high speed or bound in the Coulomb potential of point
nucleus, the variation of its mass can be ascribed to the variation of
relative ratio of hiding antimatter to matter in the particle. At two
limiting cases, the ratio approaches to 1.
\end{abstract}
The Einstein mass-energy relation $E=mc^2$ reveals the simple
proportionality between energy $E$ and mass $m$ for any matter with $c$ being
the speed of light. For a free particle moving with velocity $v$, the mass $%
m $ is related to its rest mass $m_0$ as $m=m_0(1-\frac{v^2}{c^2})^{-\frac
12}$ , which approaches infinity when $v$ approaches to $c$. On the other
hand, when an electron is bound in the Coulomb field of point nucleus with
charge $Ze$ ($e>0$) to form a hydrogenlike atom, the electron mass $m$ will
decrease. It is also known that $m$ will approach to zero when $Z$
approaches $137$. Here we show that all the above variation in particle mass
can be ascribed to the variation of relative ratio of hiding antimatter to
matter in a particle. That is the ratio of hiding positron ingredient to
electron ingredient in an electron, which determines the mass of electron.
At both sides $m\rightarrow \infty $ and $m\rightarrow 0,$ the ratio
approaches to 1.

For simplicity, we begin from a particle with mass $m_{0\text{ }}$but
without spin. It is described in nonrelativistic quantum mechanics by the
Schr\H{o}dinger equation. Then its kinetic energy reads $\frac 12m_0v^2$
with velocity $v$ being unlimited and $m_0$ unchanged. When it carries a
charge ($-e$) and is bound in the Coulomb field of a hydrogenlike atom with
potential energy 
\begin{equation}
V(r)=-\frac{Ze^2}{4\pi \varepsilon _0r}  \label{coulomb}
\end{equation}
\hspace{0in}\hspace{0in}The binding energy $B$ is well known as ($\alpha =%
\frac{e^2}{4\pi \varepsilon _0
\rlap{\protect\rule[1.1ex]{.325em}{.1ex}}h%
c}\simeq \frac 1{137}$ with $
\rlap{\protect\rule[1.1ex]{.325em}{.1ex}}h%
$ being the Plank constant) 
\begin{equation}
B=\frac{Z^2\alpha ^2}{2n^2}m_0c^2\hspace{0in}\hspace{0in}\hspace{0in}
\end{equation}
\hspace{0in}$(n=1,2,...).$ In other words, the mass of electron $m=m_0-B/c^2$
would decrease without lower bound if the charge number of nucleus $Z$ is
sufficiently large.

However, the situation becomes quite different in the theory of special
relativity. Consider a meson $\pi ^{-}$ binding in a point nucleus. Its wave
function $\Phi (z,t)$ satisfies the Klein-Gordon (K-G) equation. \cite
{Bjorken},\cite{Sakurai} 
\begin{equation}
(i
\rlap{\protect\rule[1.1ex]{.325em}{.1ex}}h%
\frac \partial {\partial t}-V(r))^2\Phi =m_0^2c^4\Phi -c^2 
\rlap{\protect\rule[1.1ex]{.325em}{.1ex}}h%
^2\nabla ^2\Phi  \label{K-G}
\end{equation}

The main point of view in this paper is as follows. We should look at $\Phi $
being composed of two kinds of fields 
\begin{eqnarray}
\theta &=&(1-\frac V{m_0c^2})\Phi +i\frac{
\rlap{\protect\rule[1.1ex]{.325em}{.1ex}}h%
}{m_0c^2}\stackrel{.}{\Phi }  \nonumber \\
\chi &=&(1+\frac V{m_0c^2})\Phi -i\frac{
\rlap{\protect\rule[1.1ex]{.325em}{.1ex}}h%
}{m_0c^2}\stackrel{.}{\Phi }  \label{two field}
\end{eqnarray}
\hspace{0in} \hspace{0in} \hspace{0in} \hspace{0in} \hspace{0in} \hspace{0in}
\hspace{0in} \hspace{0in}

Then Eq.(\ref{K-G}) can be recast into the form of coupling Schr\H{o}dinger
equations: 
\begin{eqnarray}
(i
\rlap{\protect\rule[1.1ex]{.325em}{.1ex}}h%
\frac \partial {\partial t}-V)\theta &=&m_0c^2\theta -\frac{%
\rlap{\protect\rule[1.1ex]{.325em}{.1ex}}h%
^2}{2m_0}\nabla ^2(\theta +\chi )  \nonumber \\
(i
\rlap{\protect\rule[1.1ex]{.325em}{.1ex}}h%
\frac \partial {\partial t}-V)\chi &=&-m_0c^2\chi +\frac{%
\rlap{\protect\rule[1.1ex]{.325em}{.1ex}}h%
^2}{2m_0}\nabla ^2(\chi +\theta )  \label{invariant}
\end{eqnarray}
Eq.(\ref{invariant}) is invariant under the transformation ($\overrightarrow{%
x}\rightarrow -\overrightarrow{x},t\rightarrow -t)$ and 
\begin{equation}
\hspace{0in}\theta (-\overrightarrow{x},-t)\rightarrow \chi (\overrightarrow{%
x},t),\hspace{0in}\hspace{0in}\hspace{0in}\hspace{0in}V(-\overrightarrow{x}%
,-t)\rightarrow -V(\overrightarrow{x},t)  \label{trasformation}
\end{equation}
\hspace{0in}

The meaning of $\theta $ and $\chi $ can be seen from the continuity
equation: 
\begin{equation}
\hspace{0in}\frac{\partial \rho }{\partial t}+\nabla \cdot \overrightarrow{j}%
=0
\end{equation}
with the ``probability density'' 
\begin{equation}
\rho =\left| \theta \right| ^2-\left| \chi \right| ^2=\theta ^{*}\theta
-\chi ^{*}\chi  \label{desity}
\end{equation}
and the ``current density'' 
\begin{equation}
\begin{array}{c}
\overrightarrow{j}=\frac{i
\rlap{\protect\rule[1.1ex]{.325em}{.1ex}}h%
}{2m_0}[\left( \theta \nabla \theta ^{*}-\theta ^{*}\nabla \theta \right)
+\left( \chi \nabla \chi ^{*}-\chi ^{*}\nabla \chi \right) \\ 
+(\theta \nabla \chi ^{*}-\chi ^{*}\nabla \theta )+(\chi \nabla \theta
^{*}-\theta ^{*}\nabla \chi )]
\end{array}
\end{equation}

\vspace{0in}We explain the field $\theta $ being the ``particle (matter)
ingredient'' of a particle, whereas $\chi $ being the hiding ``antiparticle
(antimatter) ingredient'' inside a particle.

See first the free motion case $V=0.$ The particle is described by a plane
wave function along $z$ axis:

\begin{equation}
\Phi \thicksim \exp \{\frac i{
\rlap{\protect\rule[1.1ex]{.325em}{.1ex}}h%
}(pz-Et)\}
\end{equation}
Beginning from $E=m_0c^2,$ $\left| \chi \right| $ increases from zero until
the limit of momentum $p\rightarrow \infty $, i.e., $E\rightarrow \infty ,$
or 
\begin{equation}
\stackunder{v\rightarrow c}{\lim }\left| \chi \right| \rightarrow \left|
\theta \right|
\end{equation}

Let us discuss the wave packet: 
\begin{equation}
\Phi (z,t)=\int_{-\infty }^\infty (\frac \sigma \pi )^{\frac 14}e^{-\frac{k^2%
}{2\sigma }}e^{i(kz-\omega t)}dk
\end{equation}
with $
\rlap{\protect\rule[1.1ex]{.325em}{.1ex}}h%
\omega =\sqrt{
\rlap{\protect\rule[1.1ex]{.325em}{.1ex}}h%
^2k^2c^2+m_0^2c^4}\backsimeq m_0c^2+\frac{
\rlap{\protect\rule[1.1ex]{.325em}{.1ex}}h%
^2k^2}{2m_0}+\cdot \cdot \cdot $

Assume $\sqrt{\sigma }\ll \frac{m_0c}{
\rlap{\protect\rule[1.1ex]{.325em}{.1ex}}h%
},$ then

\begin{equation}
\Phi \left( z,t\right) \backsimeq \frac{\left( \frac \sigma \pi \right)
^{\frac 14}}{\left( 1+\frac{i\sigma 
\rlap{\protect\rule[1.1ex]{.325em}{.1ex}}h%
t}{m_0}\right) ^{\frac 12}}\exp \{-\frac{\sigma z^2}{2\left( 1+\frac{i\sigma 
\rlap{\protect\rule[1.1ex]{.325em}{.1ex}}h%
t}{m_0}\right) ^{\frac 12}}-\frac{im_0c^2t}{
\rlap{\protect\rule[1.1ex]{.325em}{.1ex}}h%
}\}
\end{equation}

If consider $\frac{\sigma 
\rlap{\protect\rule[1.1ex]{.325em}{.1ex}}h%
t}{m_0}\ll 1$ to ignore the spreading of wave packet in low speed case ($%
v\ll c$). Then we perform a ``boost'' transformation, i.e., to push the wave
packet to high speed ($v\rightarrow c)$ case. Thus we see in the figure 1
that:

(i) The width of packet shrinks ------ Lorentz contraction.

(ii) The amplitude of $\rho $ increases------``boost'' effect.

(iii) The new observation is that both $\left| \theta \right| ^2$ and $%
\left| \chi \right| ^2$ in $\rho $ increase even more sharply while keeping $%
\left| \theta \right| >\left| \chi \right| $ to preserve $\left| \theta
+\chi \right| \thicksim \left| \Phi \right| $ invariant.

The ratio of hiding $\left| \chi \right| ^2$ to \hspace{0in}$\left| \theta
\right| ^2$ reads: 
\begin{equation}
R_{free}^{KG}=\frac{\int_{-\infty }^\infty \left| \chi \right| ^2dz}{%
\int_{-\infty }^\infty \left| \theta \right| ^2dz}=\left[ \frac{1-\sqrt{%
1-(\frac vc)^2}}{1+\sqrt{1-(\frac vc)^2}}\right] ^2  \label{R of K-G free}
\end{equation}

It is interesting to see the stationary $1S$ state (zero angular momentum
state with principal quantum number n=1) in field $V(r)$ shown in Eq.(\ref
{coulomb}).

Now the energy level is quantized to be 
\begin{equation}
E_{1S}^{KG}=m_0c^2\sqrt{\frac 12+\sqrt{\frac 14-Z^2\alpha ^2}}
\end{equation}
which is a function of Z. When $Z\rightarrow \frac 1{2\alpha }\backsimeq 
\frac{137}2$, \hspace{0in}the energy $E_{1S}$ \hspace{0in}decreases to a
lowest limit $\frac{m_0c^2}{\sqrt{2}}$. Meanwhile, the ratio 
\begin{equation}
R_{1S}^{KG}=\frac{\int \left| \chi \right| ^2d\overrightarrow{x}}{\int
\left| \theta \right| ^2d\overrightarrow{x}}=1-4\left[ 2+(y+\frac 12)^{\frac
12}+\frac{\left( y+\frac 12\right) ^{\frac 32}}{2y}\right] ^{-1}\text{%
\hspace{0in} \hspace{0in} }  \label{R of K-G 1S}
\end{equation}
(where $y=\sqrt{\frac 14-Z^2\alpha ^2,}$) increases from zero to the upper
limit 1, as shown in the figure 2.

Next turn to the electron case. Being a particle with spin $\frac 12$, it is
described by a Dirac spinor wave function

\begin{equation}
\Psi =\binom \theta \chi
\end{equation}
with four components. Here $\theta $ and $\chi ,$\hspace{0in} (each with two
components) usually called as the ``positive'' and ``negative\hspace{0in}''
energy components in the literature \cite{Bjorken},\cite{Sakurai} are just
the counterpart of $\theta $ and $\chi $ in Eqs.(\ref{two field}--\ref{R of
K-G free}) for particle without spin.

However, in this case, instead \hspace{0in}of (\ref{desity}), we have 
\begin{equation}
\rho _{Dirac}=\Psi ^{\dagger }\Psi =\theta ^{\dagger }\theta +\chi ^{\dagger
}\chi
\end{equation}

Hence for a freely moving electron wave packet, instead of figure 1, we have
figure 3. One sees that both $\theta ^{\dagger }\theta $ and $\chi ^{\dagger
}\chi $ \hspace{0in} are increasing with the velocity $v$.

But they are constrained within the boosting $\rho $ and the invariant
quantity during the boosting process is 
\begin{equation}
\overline{\Psi }\Psi =\theta ^{\dagger }\theta -\chi ^{\dagger }\chi >0
\end{equation}
where the inequality ensures that the electron is always an electron though
the hiding ``antielectron (positron)'' ingredient $\left| \chi \right| $ is
already approaching $\left| \theta \right| $ when $v\rightarrow c.$ The
ratio reads

\begin{equation}
R_{free}^{Dirac}=\frac{1-\sqrt{1-\left( \frac vc\right) ^2}}{1+\sqrt{%
1-\left( \frac vc\right) ^2}}
\end{equation}

On the other hand, when the electron is bound inside a hydrogenlike atom,
the energy level of $1S$ state is 
\begin{equation}
E_{1S}=m_0c^2[1+\frac{\alpha ^2Z^2}{\sqrt{1-Z^2\alpha ^2}}]^{-1/2}
\end{equation}
\hspace{0in} \hspace{0in} \hspace{0in} \hspace{0in} \hspace{0in} \hspace{0in}
which decreases to zero when $Z\rightarrow \frac 1\alpha \simeq 137$ as
shown in figure 4. Meanwhile the ratio

\begin{equation}
R_{1S}^{Dirac}=\frac{1-\sqrt{1-Z^2\alpha ^2}}{1+\sqrt{1-Z^2\alpha ^2}}
\end{equation}
increases from zero to $1$ , similar to Eq.(\ref{R of K-G 1S}) and the curve
in figure 2.

In summary, some discussions are in order.

(a) In nonrelativistic quantum mechanics only the particle (e.g., the
electron) is considered. The velocity of particle can enhance without a
upper limit. On the other hand, the energy of a binding particle can
decrease without a lower limit either. Its mass $m_0$ remains unchanged in
any case.

(b) In relativistic quantum mechanics, a particle is always not pure. It is
accompanied by its hiding antiparticle ingredient essentially. If a free
rest particle with mass $m_0$ described by $\theta \left( \stackrel{%
\rightarrow }{x},t\right) ,$ the accompanying $\chi \left( \stackrel{%
\rightarrow }{x},t\right) $ will be excited coherently once the particle is
set into motion or bound into an external field. Then its velocity $v$ is
bound from above by a limiting speed $c$ while its energy $E$ or its
changeable mass $m=E/c^2$ is bound from below by $m_0/\sqrt{2}$ (for $KG$
particle) or zero (for $Dirac$ particle).

(c) The common essence of any matter is the basic symmetry Eq.(\ref
{trasformation}). It could be stated as a postulate that ''the space -time
inversion $\left( \stackrel{\rightarrow }{x}\rightarrow -\stackrel{%
\rightarrow }{x},\text{ }t\rightarrow -t\right) $ is equivalent to the
transformation between particle and antiparticle''. $\cite{Ni relation}$

(d) However, inside a particle, $\theta $ always dominates $\chi $, i.e., $%
\left| \theta \right| >\left| \chi \right| $. So they do not exhibit the
symmetry Eq.(\ref{trasformation}) explicitly. Being the ``slave'' in the
particle, $\chi $ has to obey the ``master'' $\theta $. In particular, the
wave function for an electron in freely motion reads always as 
\begin{equation}
\Psi _{e^{-}}\sim \theta \sim \chi \sim exp\left\{ \frac i\hbar \left( 
\stackrel{\rightarrow }{p\cdot }\stackrel{\rightarrow }{x}-E\cdot t\right)
\right\} ,\text{ }\left( \left| \theta \right| >\left| \chi \right| \right)
\label{wave of e-}
\end{equation}

On the other hand, if we perform a space-time inversion, $\theta \left( 
\stackrel{\rightarrow }{x},t\right) \rightarrow \theta \left( -\stackrel{%
\rightarrow }{x},-t\right) =\chi _c\left( \stackrel{\rightarrow }{x}%
,t\right) $ becomes the ``master'', whereas $\chi \left( \stackrel{%
\rightarrow }{x},t\right) \rightarrow \chi \left( -\stackrel{\rightarrow }{x}%
,-t\right) =\theta _c\left( \stackrel{\rightarrow }{x},t\right) $reduces
into the "slave". Then Eq.(\ref{wave of e-}) turns into the wave function for
a positron: 
\begin{equation}
\Psi _{e^{+}}\sim \chi _c\sim \theta _c\sim exp\left\{ -\frac i\hbar \left( 
\stackrel{\rightarrow }{p}\cdot \stackrel{\rightarrow }{x}-E\cdot t\right)
\right\} ,\text{ }\left( \left| \chi _c\right| >\left| \theta _c\right|
\right)  \label{wave of e+}
\end{equation}

(e) Note that, the ``slave'' $\chi $ in a particle can not display itself as $\chi _c$ in an antiparticle, so it
does not show the opposite charge. What it can do is to pull back the motion
of $\theta $, thus the inertial mass $m$ of particle enhances without a
limit while its velocity has a limit $c$. Meanwhile, the instinct of $\chi $
demands the time evolution of phase in the wave function like that of $\chi
_c$ in Eq.(\ref{wave of e+}). But now it is forced to follow that in Eq.(\ref
{wave of e-}). They are in opposite directions. So a moving clock
accompanying the particle is slower and slower with the enhancement of $\chi 
$ inside it.

(f) The ratio $R<1$ could be viewed as an order parameter characterizing the
status of a ``particle''. Formally, if we always define $R=\int \left| \chi
\right| ^2d\stackrel{\rightarrow }{x}/\int \left| \theta \right| ^2d%
\stackrel{\rightarrow }{x}$, then $R>1$ will characterize the status of an
``antiparticle''. In other words, we look at the ``negative energy'' state
of a particle directly as the ``positive energy'' state of its antiparticle,
either for $KG$ particle or for $Dirac$ particle. It seems to us that the
historical mission of the concept of hole theory for electron is coming to
an end.

(g) Actually, all the strange effects (including the ${\it Lorentz}${\it \ }%
transformation) in special relativity can be derived by the symmetry Eq.(\ref
{trasformation}) in combination with the principles of quantum mechanics. 
\cite{Ni Chen essence}\cite{Ni Chen equation} (see also, G-j Ni and S-q
Chen, Internet, hep.th/9508069 (1995) and G-j Ni, hep.th/9708156). The
calculation shown in this paper provides the further support to the point of
view by one of us (Ni) on contemporary physics as discussed in Ref \cite{Ni
SR}\cite{Ni QM }\cite{Ni Infinity}

Acknowledgments. This work was supported in part by the NSF in China.

Correspondence should be addressed to G-j Ni.

E-mail: gjni@fudan.ac.cn

\vspace{0in}

Figure legends:

Figure 1. The wave packet of Klein-Gordon particle (e.g. $\pi ^{-}$) for
four velocities. (a) v=0.5c. (b) v=0.9c. (c) v=0.99c. (d) v=0.99999c. The 
\hspace{0in}dash , dot , and solid curves denote the profiles \hspace{0in}of 
$\left| \theta \right| ^2,$ $\left| \chi \right| ^2,$ and $\rho =$ $\left|
\theta \right| ^2-\left| \chi \right| ^2$ respectively. $\xi =m_0c(z-vt)/ 
\rlap{\protect\rule[1.1ex]{.325em}{.1ex}}h%
$ is a dimensionless quantity.

Figure 2. The dash \hspace{0in}and solid curve denote $E_{1S}^{KG}/m_0c^2$
and $R_{1S}^{KG}$ versus Z/68.5 respectively.

Figure 3. The wave packet of Dirac particle (e.g. the electron) for four
velocities. (a) v=0.5c. (b) v=0.9c. (c) v=0.99c. (d) v=0.99999c. The \hspace{%
0in}dash , dot , and solid curves denote the profiles \hspace{0in}of $\theta
^{\dagger }\theta ,$ $\chi ^{\dagger }\chi ,$ and $\rho =\theta ^{\dagger
}\theta +\chi ^{\dagger }\chi $ respectively. $\xi =m_0c(z-vt)/ 
\rlap{\protect\rule[1.1ex]{.325em}{.1ex}}h%
$ is a dimensionless quantity.

Figure 4. The dash and solid curve denote $E_{1S}^{Dirac}/m_0c^2$ and $%
R_{1S}^{Dirac}$ versus Z/137 respectively.


\begin{thebibliography}{9}
\bibitem{Bjorken}  Bjorken, J. D. and Drell, S. D.,{\it \ Relativistic
Quantum Mechanics}, McGraw-Hill Book Company, 1964.

\bibitem{Sakurai}  Sakurai, J.J.,{\it \ Advanced Quantum Mechanics, }%
Addison-Wesley, 1967.

\bibitem{Ni relation}  Ni Guang-jiong, The relation between space-time
inversion and particle-antiparticle transformation, Journal of Fudan
University (Natural Science), 1974, No.3-4, 125-134.

\bibitem{Ni Chen essence}  Ni G-j and Chen S-q, On the essence of special
relativity, ibid, 35 (3),325-334 (1996)

\bibitem{Ni Chen equation}  Ni G-j and Chen S-q, Relativistic stationary Schr%
\H{o}dinger equation for many-particle system, ibid, 36 (3) 247-252 (1997)

\bibitem{Ni  SR}  Ni G-j , To enjoy the morning flower in the
evening------Is special relativity a classical theory? Kexue (Science) 50
(1) 29-33 (1998); Internet, quant-ph/9803034.

\bibitem{Ni QM }  Ni G-j, To enjoy the morning flower in the
evening------Where is the subtlety of quantum mechanics? ibid,50 (2) 38-42
(1998); Internet, quant-ph/9804013.

\bibitem{Ni Infinity}  Ni G-j, To enjoy the morning flower in the
evening------What does the appearance of infinity in physics imply? ibid, 50
(3) 36-40 (1998); Internet, quant-ph/9806009.
\end{thebibliography}
\end{document}